\documentclass[12pt,preprint]{aastex}
\received{}
\accepted{}
\slugcomment{Accepted for AJ}             
 
\newcommand{\alp}{$\alpha\,$}                    
\newcommand{\Hp}{$\it H_{p}$}       
\newcommand{\Zs}{$Z_\odot\,$}        
\newcommand{\Ms}{$M_\odot\,$}        
\newcommand{\Rc}{$R_{core}\,$}        
\begin{document}

\title{Empirical Constraints on Convective Core Overshoot}

\author{Jong-Hak Woo and Pierre Demarque}
\affil{Department of Astronomy, Yale University, P.O. Box 208101, New Haven, CT 06520-8101 \\ jhwoo@astro.yale.edu, demarque@astro.yale.edu}

\begin{abstract}
     In stellar evolution calculations, the local pressure scale height is often 
used to empirically constrain the amount of convective core overshoot.
     However, this method brings unsatisfactory results 
for low-mass stars ($\leq$ 1.1 -1.2 \Ms for Z= \Zs) which have very small cores or 
no convective core at all.
      Following Roxburgh's integral constraint,
we implemented an upper limit of overshoot                        
within the conventional method of \alp parameterization
in order to remove an overly large overshoot effect on low-mass stars.
     The erroneously large effect of core overshoot 
due to the failure of \alp parameterization can be effectively corrected 
by limiting the amount of overshoot to $\leq$ 15 \% of the core radius. 
     15 \% of the core radius would be a proper limit of overshoot which can be 
implemented in a stellar evolution code for intermediate to low mass stars.
     The temperature structure of the overshoot region does not play a crucial role in stellar evolution since this transition region is very thin.

\end{abstract}


\section{Introduction}
Understanding the physics of convective core overshoot is important in
interpreting the color-magnitude diagrams (CMD's) and the luminosity functions of open
star clusters, and generally of intermediate-age stellar populations.
This is due to the facts that increasing the mixed core
mass modifies the evolutionary tracks of stars (affecting the isochrone
shape), and lengthens the evolutionary lifetimes of stars (affecting
age determinations and luminosity functions near the main sequence
turnoff).
  Thus, in addition to the classical problem of establishing the
chronology of star clusters in the Galaxy and nearby systems, core
overshoot affects the spectral dating of young and intermediate age galaxies
observed at large redshifts.

Convective core overshoot is loosely understood here 
as the presence of material motions and/or mixing beyond the
formal boundary for convection set by the classical Schwarzschild criterion (1906).
An early study by Saslaw \& Schwarzschild (1965), based on thermodynamic
grounds (i.e. the edge of the convective core in massive
stars is sharply defined in an entropy diagram), suggested
that little overshoot takes place at the
edge of convective cores.  For this reason, it was believed
that the presence of a gap in the CMD of
open star clusters and it's magnitude could be
used as indicators of the age and chemical
composition of the cluster (Aizenman, Demarque \& Miller 1966).
However, Shaviv \& Salpeter
(1973) pointed out that if one takes into account the
presence of hydrodynamic motions and turbulence, one might expect some
non-negligible amount of overshoot.

     In fact, observational studies of the size of gaps near
the turnoff in open star
cluster CMD's suggest better agreement with theoretical isochrones which
admit some amount of core overshoot (Maeder \& Mermilliod 1981;
Stothers \& Chin 1991; Carraro et al. 1993; Daniel et al. 1994; 
Demarque, Sarajedini \& Guo 1994;
Kozhurina-Platais et al. 1997; Nordstr\"om, Andersen \& Andersen 1997).
Several computational schemes of various degrees of
sophistication in treating the physics of overshoot have been developed for
stellar evolution codes (Prather \& Demarque 1974; Cogan 1975; Maeder 1975;
Maeder \& Meynet 1988; Bertelli et al. 1990).
    As well, studies of detached eclipsing binaries in which the components have
convective cores suggest some core overshoot of the order
of 0.2 pressure scale height (Ribas, Jordi \& Gimenez 2000).

Stothers \& Chin (1991) pointed out the high sensitivity of the amount of
convective core overshoot to the adopted opacities.  For example,
increases in radiative
opacities from the OPAL group (Iglesias \& Rogers 1996)
over the previous generation Los Alamos Opacity Library (Huebner et al. 1977),
decrease the need for
overshoot in comparison with observational data.
In fact, the nature of the overshoot depends on the details of the
local physics at
the convective core edge.  As pointed out by Zahn (1991), the
local P\'{e}clet number,
which characterizes the relative importance of radiative and turbulent
diffusivity, determines
whether
in the mixed overshoot region, penetration takes place (i.e. the temperature
gradient is
the adiabatic gradient), or rather the local radiative transfer
dominates the energy
transport (overmixing).  In the latter case, the stable temperature
gradient is unaffected by the
mixing.

Roxburgh (1989, 1992) also considered the
physics of convective core overshoot from a different point of view.
His integral constraint argument places an upper limit on the amount
of core overshoot which can take place in the stellar interior.
Zahn (1991) argued that overshoot is practically adiabatic
at the edge of the convective core, except in a thin transition layer,
and therefore the Roxburgh's integral constraint defines the
convective core size.

More information can be gained from numerical simulations of convection.
Because of the long relaxation times involved, it is not possible to
perform 3D simulations of convective cores in full physical detail.  Useful
scaling information can however be derived from idealized 3D simulations
(Singh, Roxburgh \& Chan 1995, 1998).

The likely presence of differential rotation deep 
within stars brings further complexity to the
overshoot problem, as one might expect a shear layer to develop
at the edge of a convective core
(Pinsonneault et al. 1991).  Deupree (1998, 2000) considered
the combined effects of rotation and convective overshoot in massive stars,
which have large convective cores, with the help of 2D hydrodynamic calculations.
Rotation likely plays a role in the case of less massive stars as well 
(in the range 1.5-2.0 \Ms), with
shear driven turbulence near the edge of the convective core inducing
mixing from the helium enriched core into the envelope.

Since a complete physical theory predicting the amount of overshoot for a
star of a given mass and chemical composition, free
of arbitrary parameters, is
not available at this point,  we approach the
problem of core overshoot in the spirit
of the work of Rosvick \& VandenBerg
(1998) in this paper. We consider and test parameterizations of convective
overshoot, which are
compatible with sound physical principles and with the growing data
available on open cluster CMDs.  These
simple parameterizations have the merit of being readily implemented
in a stellar evolution code.

\section{Test of Overshoot Treatments}
 
     The local pressure scale height is often used to empirically constrain 
the amount of convective core overshoot in stellar evolution calculations.
     This conventional method assumes that 
the radial size of the overshoot region is proportional to the pressure scale 
height at the edge of a convective core, i.e. $d_{os}$ =\alp \Hp. 

     Observational studies of open clusters show that a moderate amount of 
core overshoot is essential to explain the observed gap 
near the turnoff in CMDs. 
     In a study of NGC 2420, 
Demarque et al. (1994) found that \alp =0.23 is required for the best
fit CMD. Also, Kozhurina-Platais et al. (1997) estimated \alp =0.20 and 0.25 for NGC 3680 and NGC 752 repectively by isochrone fitting (cf. Nordstr\"om et al. 1997 for NGC 3680).
Overshoot effects appear to decrease due to the decreasing mass of stars  
near the turnoff in older clusters (Carraro et al. 1994; Sarajedini et al. 1999). 
     Although open clusters do not have many member stars, these studies
of near solar metallicity clusters
show that the conventional method constrains the amount of core 
overshoot relatively well for intermediate-mass stars.

     However, this pressure scale height method brings unsatisfactory results 
for low-mass stars ($\lesssim$ 1.1- 1.2 \Ms for Z= \Zs). These stars have very small cores or 
no convective core at all, thus the local pressure scale height at the convective core edge 
would be very large implying a large amount of overshoot.
     Figure 1 illustrates the effects of the convective core and core overshoot 
on stellar evolution. 
With no overshoot, models of mass $\ge$ 1.2 \Ms show a MS hook which is caused by abrupt contraction of a convective core due to the flat hydrogen profile. 
     When overshoot is included, evolutionary tracks show a redder MS hook 
due to the hydrogen supply from mixing in the overshoot region which lengthens 
the hydrogen core burning lifetime. 
     However, the effect of overshoot on low-mass
stars seems too large compared with observations. If 
the amount of overshoot is set to 0.2 \Hp, low-mass stars 
with 1.0 and 1.1 \Ms show a prominent overshoot effect in their evolutionary shape. 
However, these low-mass stars are not expected to have large overshoot regions 
since relatively old open clusters do not show an overshoot gap near the turnoff.
     Therefore, simple use of the \alp parameter cannot give 
consistent results for a wide range of masses and ages.

    Roxburgh's integral constraint (Roxburgh 1989; Zahn 1991) is 
another way to quantify the amount of core overshoot
since it provides an upper limit to the extent of convective penetration in the following
formula:

\begin{equation}
\int^{r_{c}}_0 (L_{rad} - L_{tot}) \frac{1}{T^2} \frac{dT}{dr} dr = 0
\end{equation}

    where $\it L_{tot}$ is the total luminosity produced by nuclear reactions, 
$\it L_{rad}$ is the radiative luminosity, and $r_{c}$ is the radius of the effective core.
Since the energy is transported by both radiation and convection inside a convective core,
the integrand is positive until it reaches the Schwarzschild boundary 
where $\it L_{tot}$ = $\it L_{rad}$. 
    The region of convective overshoot is located beyond this boundary 
where $\it L_{rad}$ $>$ $\it L_{tot}$ 
up to the point $\it r$ = $r_{c}$, which satisfies the constraint.

    However, Canuto (1997) pointed out in his theoretical investigation that 
Roxburgh's integral only gives an upper limit to the overshoot
extent.
Rosvick \& VandenBerg (1998) used this constraint on NGC 6819 
and found that a factor of 0.5 in Roxburgh's integral is required for the best
fit CMD. Therefore, simply adopting Roxburgh's integral 
to quantify the amount of overshoot is not valid, 
and the fuzzy factor (analogous to the \alp parameter in the pressure scale height method)
should be tested for clusters of various ages and metallicities. 
 
    Faced with the current situation of the core overshoot treatments, 
we developed a consistent approach which can be used regardless of stellar mass. 
      Our simple approach of core overshoot is discussed in the next section.
All stellar evolutionary tracks are constructed with Yale Rotating Evolution Code (YREC) with OPAL opacities (Iglesias \& Rogers 1996). The general description of stellar evolution 
models can be found in Yi et al. (2001). 

\subsection{Maximum amount of core overshoot ($\beta$ limit)}

    Roxburgh (1992) showed that his integral constraint 
gives an upper limit on the extent of convective penetration 
($d_{max}\,$ =$\beta$ \Rc).
For very small convective cores, the maximum amount of 
the extent of penetration is $\sim$ 18 $\%$ of 
the core radius, independent of the details of energy generation and opacity. 
The maximum amount, $\beta$ limit, varies from 0.18 to 0.4 depending on the size of the convective core.
 
      Following Roxburgh's integral constraint,
we implemented an upper limit of overshoot 
within the conventional pressure scale height method 
in order to remove the erroneously large overshoot effect on low-mass stars.
      In other words, we limited the core overshoot in a way 
that the amount of overshoot, $d_{os}\,$ (= \alp \Hp ), cannot be larger 
than a portion of the core radius, $d_{max}\,$ (= $\beta$ \Rc ). 
       The merit of this modification is that the ad hoc size of overshoot regions
for low-mass ($\lesssim$ 1 \Ms) stars can be effectively corrected
without affecting the evolution of higher mass stars.

        Since the integral constraint is obtained for convective penetration 
which maintains a nearly adiabatic temperature structure,
we used the adiabatic temperature gradient in the overshoot region. 
        The effect of the temperature structure will be discussed in the following section.
        Accepting the result of open cluster studies, 
we used \alp =0.2, and performed numerical experiments with various 
$\beta$ limits for solar metallicity stars.

       The effect of the $\beta$ limit on stellar evolution is presented in Fig. 2. 
For low-mass stars, the presence of the overshoot effect is very sensitive to 
the choice of the upper limit of overshoot.
       It can be easily noticed that the overshoot effect on 
low-mass stars (mass $\lesssim$ 1.1 \Ms) becomes negligible for $\beta$ $\leq$ 0.15. 
       However, if the upper limit is greater than 20 \% of the core radius, 
the overshoot effect is still too large for low-mass stars.
       This result is consitent with Roxburgh's value of 0.18 
for a nearly zero convective core radius.
       Thus, the erroneously large effect of core overshoot due to the failure of \alp parameterization for low-mass stars can be effectively corrected by limiting the amount of overshoot to $\lesssim$ 15 \% of the core radius. 

        The choice of $\beta$ does not significantly affect the evolution of intermediate-mass stars 
(1.5 -  2.0 \Ms) when $\beta$ $\geq$ 0.15 is used. These models are consistent with those of the same \alp without limiting the overshoot extent.
        However, if the amount of overshoot is 
restricted to much less than 15 \% of the core radius,  the overall overshoot effect 
would be significantly reduced.
        Therefore $\beta$ $\ll$ 0.15 for intermediate-mass stars gives 
effectively the same result with a smaller \alp of the conventional method, 
which is undesirable in our effort to modify the \alp paramerterizaion with the $\beta$ limit
for a given \alp value. 
        Combining the criteria for low-mass and intermediate-mass stars, 
we conclude that 15 \% of the core radius ($\beta$ =0.15) is a proper 
limit of overshoot for intermediate to low-mass stars.

        Due to the effects of overshoot on stellar evolution, such as shape of evolutionary locus 
near turnoff and time scale in hydrogen-burning stage, the ensuing isochrones are 
expected to show a different shape. We used 1.0 - 2.0 \Ms evolutionary tracks with 
mass bin size 0.1 \Ms in order to generate isochrones and test the effect of the $\beta$ limit. 
        We compared \alp =0.2 isochrones with/without limiting the amount of overshoot (Fig. 3). 
        As expected from the comparison of evolutionary tracks, 
$\beta$ =0.2 isochrones are not much different from the isochrones without the $\beta$ limit.
        When overshoot is limited to 10 \% of the core radius, overall 
isochrones are dimmed due to the reduced size of the overshoot region,
especially for younger ages. This result is similar to when smaller \alp is used. 
        Thus, $\beta$ =0.1 is too small to keep the amount of overshoot of intermediate-mass stars given by \alp =0.2.

      Focusing on the modification of the \alp parameter method, 
we increased $\beta$ with increasing stellar mass in order to minimize the effect of limiting 
the extent of overshoot for stars of mass $\gtrsim$ 1.5 \Ms.
      Since the maximum extent of core overshoot depends on the size of a 
convective core, higher mass stars would have larger limits according to 
Roxburgh's integral constraint. 
      Thus, we assumed a simple linear relation between
stellar mass and the upper limit of overshoot 
($\beta = aM+b$). The coefficients are determined 
for $\beta$ = 0.1 and 0.4 for 1 and 2 \Ms stars respectively. 
       Fig. 4 demonstrates how the mass-$\beta$ relation changes stellar evolution.
Compared with $\beta$=0.15 models, models with increasing $\beta$ show slightly 
bluer and brighter isochrones for younger ages. 
      Although increased $\beta$ models are closer to the models without the $\beta$ limit, the difference is not significant.

\subsection{Penetration and Overmixing}

    The overshoot region generally represents a well mixed zone 
between convective and radiative zones. 
    The temperature structure of the overshoot region 
depends on the physical conditions 
such as the ratio of radiative to kinetic 
timescales at the convective core boundary.
    According to Zahn (1991; also see Demarque et al. 1994),
if the adiabatic cells penetrate into 
the radiative zone sufficiently far, then the temperature gradient
of the overshooting zone would be adiabatic also (convective penetration).
   Or, if the adiabatic cells dissolve rapidly, then only
mixing would occur with the remaining radiative gradient 
(overmixing).

    However, it is not well known whether the temperature structure of 
the overshoot region is adiabatic or radiative.
    In order to test the effects of the temperature structure on stellar evolution,
we generated evolutionary tracks and isochrones with/without
forcing the temperature gradient of the overshoot region to be adiabatic. 
  
    Stellar evolutionary tracks of solar metallicity show almost no difference 
between overmixing and convective penetration when the conventional method of \alp =0.1 - 0.3 is used.
    This is also true when the amount of overshoot is limited 
by any factor of the core radius (Fig. 5).
    We conclude that temperature structure of the overshoot region does not play a crucial role in stellar evolution,
at least when the local pressure scale height method is used.
    This is due to the fact that this transition region is very thin.
Note that the effect is slightly larger in low-mass stars.

\section{Conclusion}
    This paper presents an approach to the treatment
of convective core overshoot in stellar
evolutionary models for stars with convective cores.
    By limiting the amount of convective core overshoot, the failure of 
the pressure scale height method can be effectively corrected. 
    15 \% of the core radius would be a proper limit of core overshoot
for intermediate to low mass stars.                     
     We intend to use evolutionary tracks based on this
approach to explore population synthesis and the spectral
dating of high redshift galaxies.

    It is crucial to test stellar evolution by observation.
    Several observational approaches to test stellar models with convective
overshoot are under way.      
    The systematic study of open star clusters that exhibit
hydrogen exhaustion phases provides
important constraints that stellar models must satisfy.
The WIYN Open Cluster Study (WOCS)
project, provides a comprehensive study of open clusters in the Galaxy
selected to provide a range
of chemical compositions and ages.  In addition to photometric and
spectroscopic observations of
individual stars, WOCS includes an astrometric proper motion
study to establish cluster
membership, a necessary condition.  Because the number of stars in open
clusters is small, establishing membership is essential for
carrying out detailed tests of stellar evolutionary models
using a cluster CMD and luminosity function.  This
advantage is demonstrated in recent deep proper motion studies
of NGC 188 (Dinescu et al. 1996) and NGC 3680 (Kozhurina-Platais et al.
1997).  An
excellent overview of WOCS objectives and results obtained so
far has been written by Mathieu (2000).

Observations of the CMDs of some of
the more populous star clusters in the Magellanic Clouds will
provide more information on core overshoot in intermediate age
stellar populations.  Although more distant than open clusters
in the Galaxy, the Magellanic Cloud clusters have the advantage
of testing the evolution of stars with a lower metallicity.

Finally, stellar seismology, by analyzing oscillations that
propagate deep in stellar interiors, promises to probe
sensitively the convective cores of stars evolving off
the main sequence in the intermediate mass
range.  Similarly, the mass of the exhausted helium
core in subgiants and giants can be derived on the basis
of physical models.  Already observations of oscillations
of nearby stars near the main
sequence and in the subgiant and giant evolutionary phases
have been done from the ground
(Kjeldsen et al. 1995; Martic et al. 1999; Bedding et al. 2001)
and in space (Buzasi et al. 2000).
Preliminary analysis of some of these observations with the
help of stellar models show the
unmistakable signature (mixed p- and g-modes) of exhausted
core size in the case of the more
evolved stars (Christensen-Dalsgaard, Bedding \& Kjeldsen 1995;
  Guenther \& Demarque 1996).

\acknowledgements
Acknowledgement.  This research has been supported in part by
NASA grant NAG5-8406.

\clearpage

\begin{figure}
\plotone{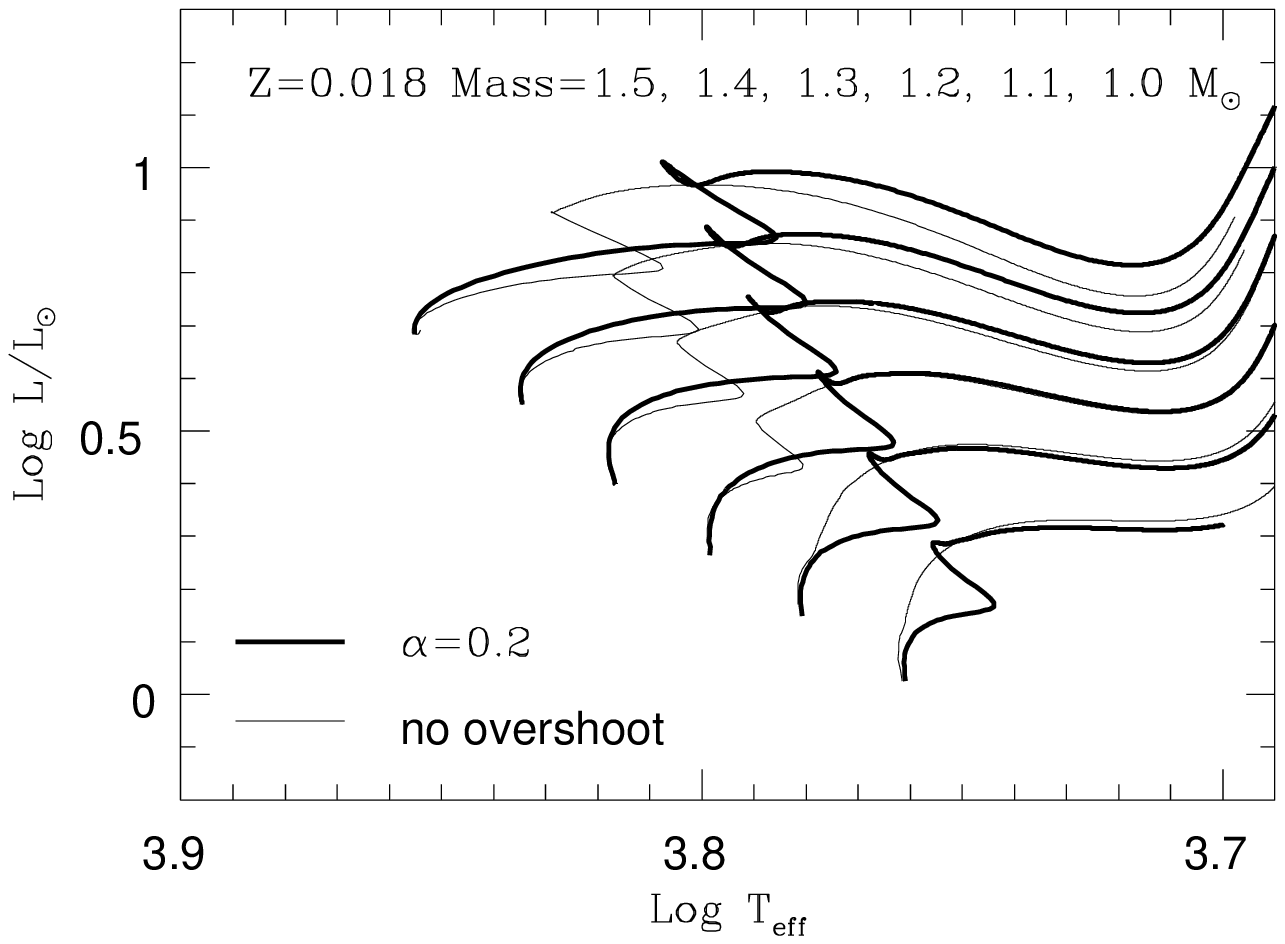}
\caption{The effect of convective core overshoot on stellar evolution 
for solar metallicity stars of 1.5 to 1.0 \Ms . 
Overshoot tracks have a redder MS hook due to the lengthened H core burning lifetime.
The MS hook of low-mass stars in overshoot models show that the 
\alp  parameter method fails due to the very large pressure 
scale height at the edge of the small convective core. \label{fig1}}
\end{figure}   

\clearpage

\begin{figure}
\plotone{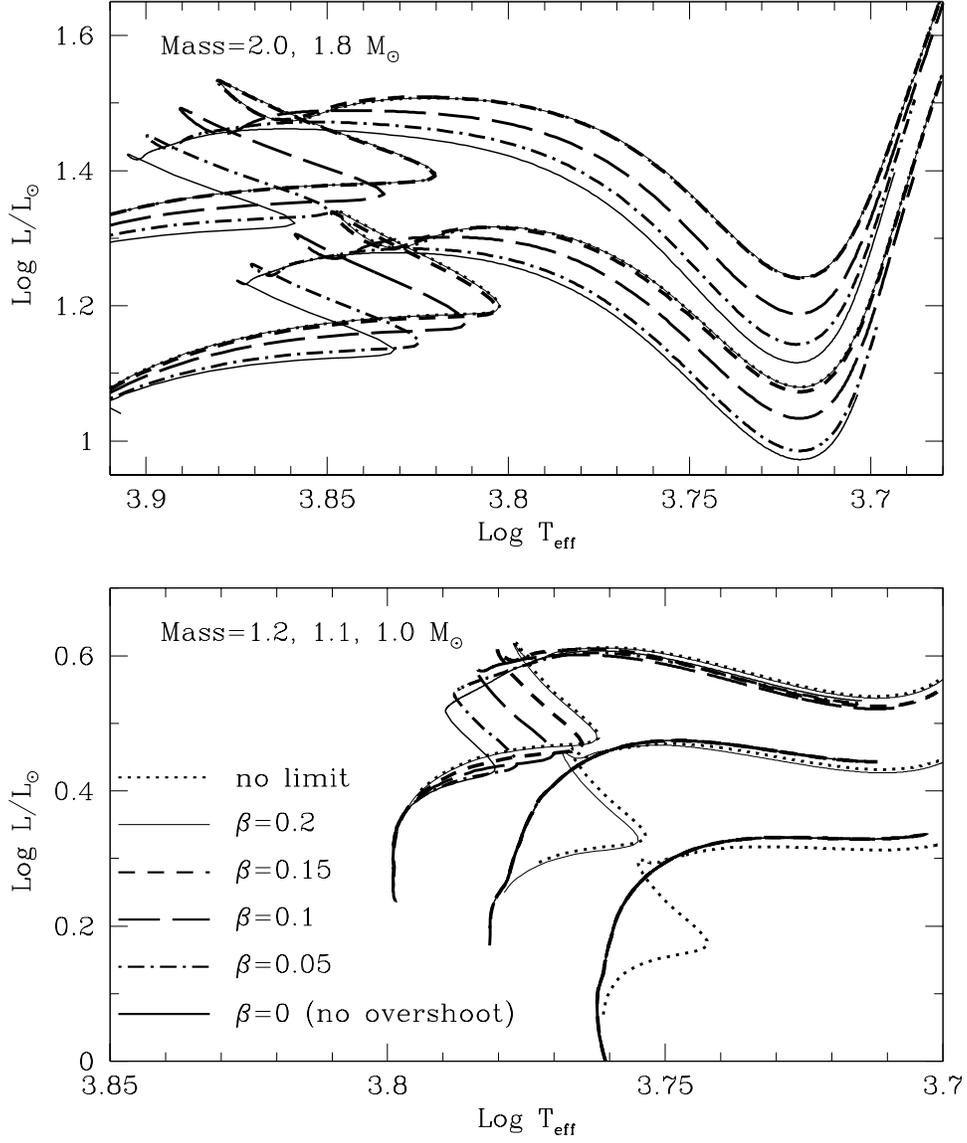}
\caption{The effect of $\beta$ limit on stellar evoluton for Z=0.018. 
The amount of core overshoot is limited to $\beta$ factor of \Rc. 
Note that 
models with $\beta$ $\geq$ 0.15 show nearly identical tracks for intermediate-mass stars (top panel).
However, the overshoot effect is not present for low-mass stars when
$\beta$ $\leq$ 0.15 (bottom panel),
which suggests that the erroneously large effect of overshoot on low-mass stars can be corrected by limiting the extent of core overshoot. 
\label{fig2}} 
\end{figure}   

\clearpage

\begin{figure}
\plotone{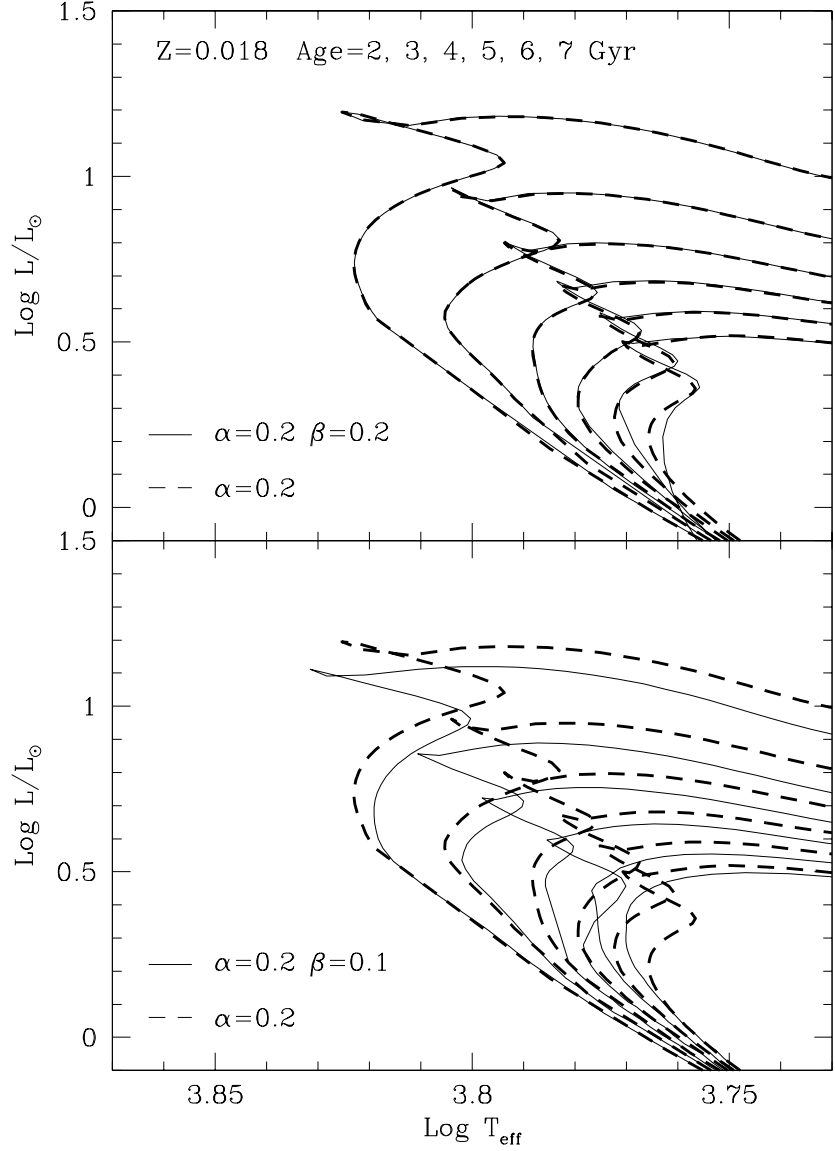}
\caption{The effect of the $\beta$ limit on the isochrones of Z=0.018. 
$\beta$=0.2 models are not much different from the models without limiting core overshoot, and repeat the failure of \alp parameterization (top panel).
When overshoot is limited to 10 $\%$ of the core radius, 
overall isochrones are dimmed, which is similar to the result of smaller \alp (bottom panel). \label{fig3}}
\end{figure}   

\clearpage

\begin{figure}
\plotone{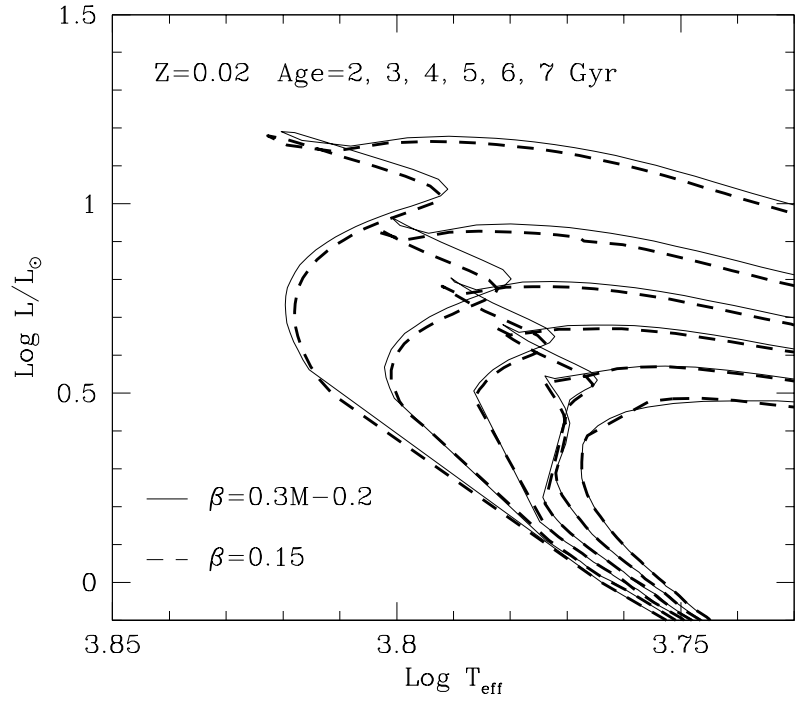}
\caption{$\beta$ limit as a function of the stellar mass.
$\beta$ is set up to be 0.4 for 2 \Ms and 0.1 for 1 \Ms stars respectively. 
Compared with constant $\beta$ models, models with increasing $\beta$ show slightly brighter isochrones for younger ages. However, the difference is not siginificant. \label{fig4}}        
\end{figure}   

\clearpage

\begin{figure}
\plotone{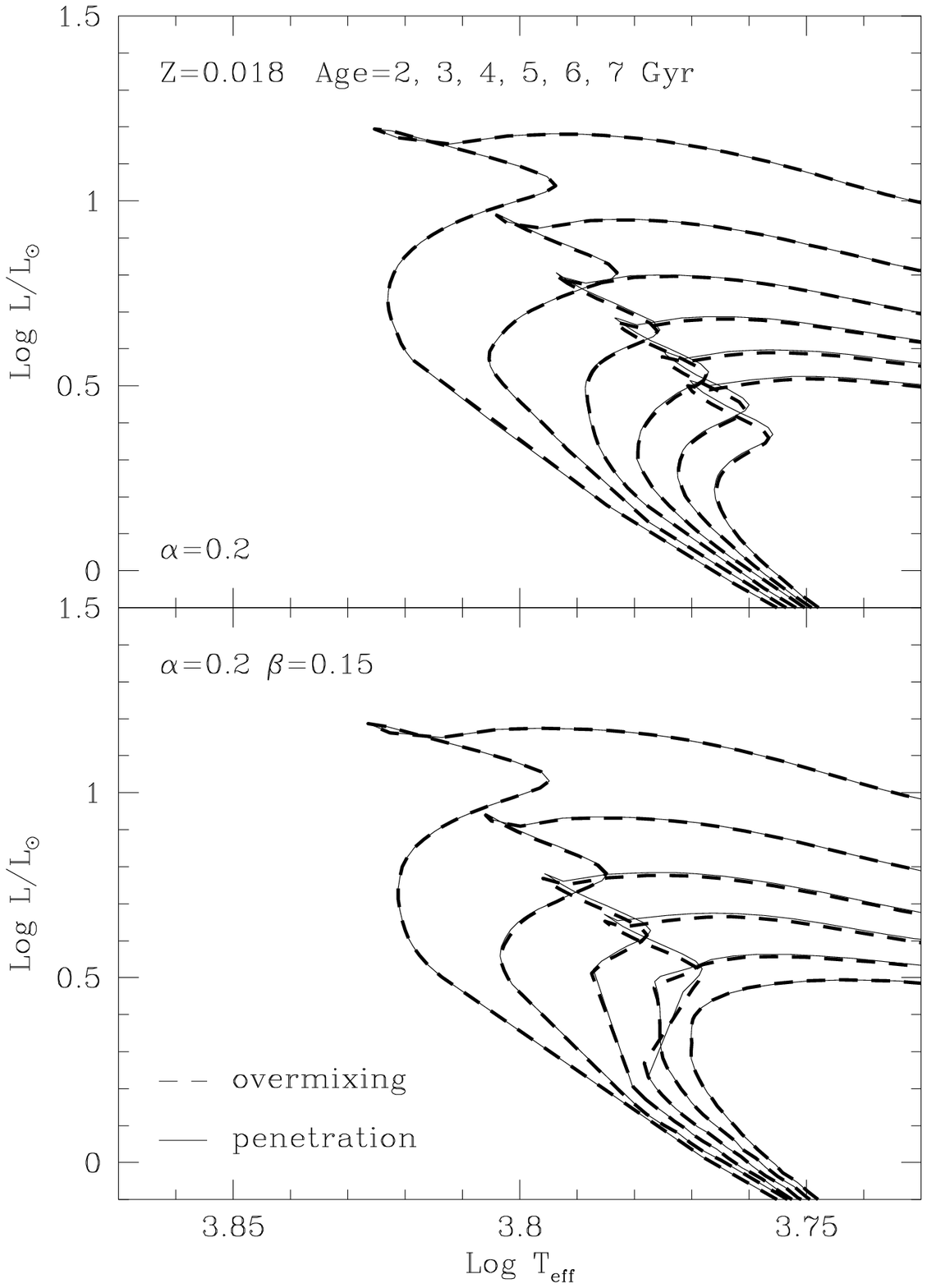}
\caption{The effect of the temperature structure of the overshoot region.          
The temperature structure of the overshoot region does not affect stellar evolution significantly since this transition region is very thin. \label{fig5}}
\end{figure}   


\begin{thebibliography}{}
\bibitem[Audard et al. (1995)]{aud}Audard, N., Provost, J., \& Christensen-Dalsgaard, J. 1995, A\&A, 297, 427
\bibitem[Bedding et al. (2001)]{bed}Bedding, T.R., Butler, R.P., Kjeldsen, H., Baldry, I.K., O'Toole, S.J., Tinney, C.G., Marcy, G.W., Kienzle, F. \& Carrier, F. 2001, ApJL accepted (astro-ph/0012417)
\bibitem[Bertelli et al. (1990)]{ber}Bertelli, G., Betto, R., Chiosi, C., Bressan, A., Nasi, E. 1990, A\&AS, 85, 845
\bibitem[Buzasi et al. (2000)]{buzasi}Buzasi, D., Catanzarite, J., Laher, R., Conrow, T., Shupe, D., Gautier, T.N. III, Kreidl, T. \& Everett, D. 2000, \apj, 532, 133
\bibitem[Canuto (1997)]{canuto}Canuto, V. M. 1997, ApJ, 489, L71
\bibitem[Carraro et al. (1993)]{carraro}Carraro, G., Bertelli, G., Bressan, A. \& Chiosi, C. 1993, A\&AS, 101, 381
\bibitem[Carraro et al. (1994)]{carraro2}Carraro, G., Chiosi, C., Bressan, A. \& Bertelli, G. 1994, A\&AS, 103, 375
\bibitem[Christensen-Dalsgaard et al. (1995)]{chris}Christensen-Dalsgaard, J., Bedding, T.R. \& Kjeldsen, H. 1995, ApJ, 443, 29
\bibitem[Cogan (1975)]{cogan}Cogan, B.C. 1975, ApJ, 201, 637
\bibitem[Daniel et al. (1997)]{daniel}Daniel, S. A., Latham, D. W., Mathieu, R. D. \& Twarog, B. A. 1997, PASP, 106, 281
\bibitem[Demarque et al. (1994)]{de94} Demarque, P., Sarajedini, A., \& Guo, X. J. 1994, \apj, 426, 165
\bibitem[Deupree (1998)]{deu}Deupree, R.G. 1998, ApJ, 499, 340
\bibitem[Deupree (2000)]{deu2}Deupree, R.G. 2000, ApJ, 543, 395
\bibitem[Dinescu et al. (1996)]{din}Dinescu, D.I., Girard, T.M., van Altena, W.F., Yang, T.-G. \& Lee, Y.-W. 1996, AJ, 111, 1205
\bibitem[Gunther (1991)]{guen}Guenther, D.B. 1991, ApJ, 375, 352
\bibitem[Gunther and Demarque (1996)]{GD}Guenther, D.B. \& Demarque, P. 1996, ApJ, 456, 798
\bibitem[Gunther et al. (2000)]{GDetal}Guenther, D.B., Demarque, P., Buzasi, D., Catanzarite, J., Laher, R., Conrow, T. \& Kreidl, T. 2000, ApJ, 530, 45
\bibitem[Huebner et al. (1977)]{hue}Huebner, W.F., Merts, A.L., Magee, N.H., \& Argo, M.F. 1977, Los Alamos Opacity Library, Los Alamos Scientific Laboratory Rep. No. LA-6760-M
\bibitem[Iglesias and Rogers (1996)]{IR}Iglesias, C.A. \& Rogers, F.J. 1996, ApJ, 464, 943
\bibitem[Kjeldsen et al. (1995)]{Kje}Kjeldsen, H., Bedding, T.R., Viskum, M. \& Frandsen, S. 1995, AJ, 109, 1313
\bibitem[Kozhurina-Platais et al. (1997)]{koz} Kozhurina-Platais, V., Demarque, P., Platais, I., Orosz, J. A., \& Barnes, S. 1997, \aj, 113, 1045
\bibitem[Meader (1975)]{meader}Maeder, A. 1975, A\&A, 43, 61
\bibitem[Meader and Mermilliod (1981)]{MM}Maeder, A. \& Mermilliod, J.-C. 1981, A\&A, 93, 136
\bibitem[Meader and Meynet (1988)]{MM2}Maeder, A. \& Meynet, G. 1988, A\&AS, 76, 411
\bibitem[Martic et al. (1999)]{metal}Martic, M., Schmitt, J., Lebrun, J.-C., Barban, C., Connes, P., Bouchy, F., Michel, E., Baglin, A., Appourchaux, T. \& Bertaux, J.-L. 1999, A\&A, 351, 993
\bibitem[Mathieu (2000)]{mathieu}Mathieu, R.D. 2000, in Stellar Clusters and Associations: Convection, Rotation, and Dynamos, ASP Conf. Ser. Vol, 198, eds. R. Pallavicini, G. Micela, and S. Sciortino, p. 517
\bibitem[Nordst\"om et al. (1997)]{nor}Nordstr\"om, B., Andersen, J. \& Andersen, M. I. 1997, A\&A, 322, 460 
\bibitem[Pinsonneault et al. (1991)]{pin}Pinsonneault, M.H., Deliyannis, C.P. \& Demarque, P. 1991, ApJ, 367, 239
\bibitem[Prather and Demarque (1974)]{prat}Prather, M.J. \& Demarque, P. 1974, ApJ, 193, 109
\bibitem[Ribas et al. (2000)]{rib}Ribas, I., Jordi, C., Gimenez, A. 2000, MNRAS, 318, 55
\bibitem[Roxburgh (1989)]{rox1} Roxburgh, I. W. 1989, A\&A, 211, 361
\bibitem[Roxburgh (1992)]{rox2} Roxburgh, I. W. 1992, A\&A, 266, 291
\bibitem[Rosvick and VandenBerg (1998)]{rosvick} Rosvick, J. M. \& VandenBerg, D. A. 1998, \aj, 115, 1516
\bibitem[Sarajedini et al. (1999)]{sara} Sarajedini, A., von Hippel, T., Kozhurina-Platais, V., \& Demarque, P. 1999 \aj, 118, 2894
\bibitem[Saslaw and Schwarzschild (1965)]{sch}Saslaw, W.C. \& Schwarzschild, M. 1965, ApJ, 142, 1468
\bibitem[Schwarzschild (1906)]{sch2}Schwarzschild, K. 1906, G\"{o}ttingen Nach., 195, 41
\bibitem[Salpeter (1973)]{shaviv}Shaviv, G. \& Salpeter, E.E. 1973, ApJ, 184, 191
\bibitem[Singh et al. (1995)]{sing}Singh, H.P., Roxburgh, I.W. \& Chan, K.L. 1995, A\&A, 295, 703
\bibitem[Singh et al. (1998)]{sing2}Singh, H.P., Roxburgh, I.W. \& Chan, K.L. 1998, Ap\&SS, 261, 53
\bibitem[Stothers and Chin (1991)]{stot}Stothers, R. B. \& Chin, C.-W. 1991, ApJ, 381, L67
\bibitem[Yi et al. (2001)]{yi}Yi, S., Demarque, P., Kim, Y.-C., Lee, Y.-W., Ree, C. H., Lejeune, T., \& Barnes, S. 2001, ApJS, accepted (astro-ph/0104292)
\bibitem[Zahn (1991)]{zahn} Zahn, J.-P. 1991 A\&A, 252, 179
\end{thebibliography}
\end{document}